\documentclass[11pt]{article}
\usepackage{moriond,epsfig}
\usepackage{amssymb,amsmath}

\def\Journal#1#2#3#4{{#1} {\bf #2}, #3 (#4)}

\def\PLB{{\em Phys. Lett.}  B}
\def\PRL{\em Phys. Rev. Lett.}

\def\EPJ{{\em Eur. Phys. J.} C}

\def\be{\begin{equation}}
\def\ee{\end{equation}}
\def\bea{\begin{eqnarray}}
\def\eea{\end{eqnarray}}

\begin{document}
\vspace*{4cm}
%\title{Charmless two-dody $B$-hadron decays at LHCb with 2010 data}
\title{Measurements of $A_{CP}(B^0\rightarrow K^+ \pi^-)$ and
  $A_{CP}(B_s\rightarrow \pi^+ K^-)$ at LHCb}
\author{Stefano Perazzini\\ on behalf of the LHCb Collaboration }
\address{Dipartimento di Fisica, Via Irnerio 46, \\
40126 Bologna, Italy}

%\address{Department of Physics, Theoretical Physics, 1 Keble Road,\\
%Oxford OX1 3NP, England}

\maketitle\abstracts{\tolerance=0
  The LHCb experiment is designed to perform flavour physics measurements at the Large Hadron Collider.
  Using data collected during the 2010 run, we reconstruct a sample of
  $H_b\rightarrow h^+h'^-$ decays, where $H_b$ can be
  either a $B^0$ meson, 
  a $B_s^0$ meson or a $\Lambda_b$ baryon, while
  $h$ and $h^\prime$ stand for $\pi$, $K$
  or $p$. 
  We provide preliminary values of the direct $\mathcal{CP}$ asymmetries of the
  neutral $B^0$ and $B_s^0$ mesons
  $A_{CP} (B^0\rightarrow K^+\pi^-) = -0.074 \pm 0.033\mathrm{(stat.)}
  \pm 0.008\mathrm{(syst.)}$ 
  and $A_{CP}(B_s^0\rightarrow\pi^+K^-)=0.15 \pm
  0.19\mathrm{(stat.)} \pm 0.02\mathrm{(syst.)}$.}

\section{Introduction}\label{sec:intro}
The family of $H_b\rightarrow h^+h'^-$ comprises a large set
of decays, namely: $B^0\rightarrow \pi^+ \pi^-$,
$B^0\rightarrow K^+ \pi^-$, 
$B_s^0\rightarrow K^+ K^-$, 
$B_s^0\rightarrow \pi^+ K^-$, 
$\Lambda_b\rightarrow p K^-$,
$\Lambda_b\rightarrow p \pi^-$,
$B^0\rightarrow K^+ K^-$,
$B_s^0\rightarrow \pi^+ \pi^-$ plus their $\mathcal{CP}$-conjugate
states. Such decays are matter of great interest, as they are  sensitive
probes of the Cabibbo-Kobayashi-Maskawa~\cite{cab,koba} matrix 
and have the potential to reveal the presence of New
Physics (NP)~\cite{fle,flei}.
%The LHCb experiment has an excellent potential to dramatically improve the world
%knowledge of such decays. In this note, we present preliminary values of the direct CP
%asymmetries in the B0 → K+π− and Bs0 → π+K− decays, defined in terms of decay rates
%In the Standard Model (SM), the amplitudes of such decays receive
%contributions both from tree and from penguin topologies. In addition one have to
%take into account the oscillation of $\mathrm{B^0}$ and
%$\mathrm{B_s^0}$ mesons between their CP-conjugate states, process
%governed by box diagrams.\\
%The presence of loops and box diagrams, respectively in the decay graphs
%and in the mixing graphs, makes this analysis a suitable place where
%to look for New Physics effect\cite{fle,flei}. 
NP may alter in a subtle but sizeable way the Standard Model (SM) prediction
of the $\mathcal{CP}$ asymmetries in these decays.
%In the end, combining the time dependent $CP$-asymmetries of
%$B^0\rightarrow \pi^+\pi^-$ and $B_s^0\rightarrow K^+K^-$, and within
%the validity of the U-spin symmetry (invariance of strong interaction 
%dynamics under the $d \leftrightarrow s$ quarks
%exchange), it will be possible to extract the angle $\gamma$ of the
%Cabibbo-Kobayashi-Maskawa (CKM) matrix~\cite{cab,koba}. %A different
%measured value of angle $\gamma$ from these decays with respect to
%that extracted from tree-level decays might reveal the presence of New
%Physics inside the already mentioned loop diagrams.\\
In the following, we will present the preliminary measurements of
the direct $\mathcal{CP}$ asymmetries in the $B^0\rightarrow K^+ \pi^-$ and
$B_s^0\rightarrow \pi^+ K^-$ decays, obtained using the data
collected by LHCb during the 2010 at a centre of mass energy of $7$~TeV, corresponding to 
an integrated luminosity of $\int\mathcal{L}\mathrm{d}t\simeq 37$~$\mathrm{pb}^{-1}$.
Such direct $\mathcal{CP}$ asymmetries are
defined in terms of decay rates of $B$-hadrons as
$A_{CP} = \left[\Gamma\left(\bar{B}\rightarrow\bar{f}\right) - \Gamma\left( B\rightarrow f\right)\right]/
\left[\Gamma\left(\bar{B}\rightarrow\bar{f}\right)+\Gamma\left(
    B\rightarrow f\right)\right]$.

\section{The LHCb detector}\label{sec:detector}
The LHCb detector~\cite{jinst} is a single arm spectrometer in the forward direction.
%covering a pseudo-rapidity range $1.8 < \eta < 4.9$. 
It is composed of a vertex detector around the interaction region, a set of tracking
stations in front of and behind a dipole magnet that provides a field
integral of $4$~Tm, two Ring-Imaging Cherenkov (RICH) detectors,
electromagnetic and hadronic calorimeters complemented with pre-shower
and scintillating pad detectors, and a set of muon chambers. The two
RICH detectors are of particular importance for this analysis, as they provide the particle
identification (PID) information needed to disentangle the various 
$H_b\rightarrow h^+h'^-$ final states. They are able to
efficiently separate $\pi$, $K$ and protons in a momentum range from $2$~GeV/c
up to and beyond $100$~GeV/c. RICH-1 is installed in front of the magnet and uses
Areogel and $\mathrm{C_4F_{10}}$ as radiators, while RICH-2 is
installed behind the magnet and employs $\mathrm{CF_4}$.

\section{Events selection}
%The online selection of the events used in this analysis is performed
%by the two-level trigger system of LHCb. The first level (Level 0) is a
%hardware trigger implemented with custom electronic boards.
%In particular, the $H_b\rightarrow h^+ h'^-$ decays are
%principally selected by the hadronic Level 0 trigger, that looks for
%high transverse energy clusters in the hadronic calorimeter, as
%signatures of high transverse momentum hadrons. The total Level 0
%trigger output rate is $1.1$~MHz.\\
%The second level trigger, so-called High Level Trigger (HLT), is
%software-based and performs a full reconstruction of events. The HLT
%strategy is to look for events with at least one track characterized
%by a large trasnverse momentum and a large impact parameter with
%respect to all the primary vertices in the event. The goal of this software trigger is to
%reduce the recorded data to a rate of about $2$~kHz.\\
The $H_b\rightarrow h^+h'^-$ decays are principally selected by the
two-level hadronic trigger of LHCb. The first level (Level 0) is based
on custom electronic boards, selecting events with high transverse
energy clusters in the hadronic calorimeter. The second level, so
called High Level Trigger (HLT), is software-based and selects events
with at least one track with high transverse momentum and large impact
parameter with respect to all reconstructed primary vertices.\\ 
The events used in this analysis are extracted from the triggered data
using two different offline selections, each one targeted to achieve the best
sensitivity on $A_{CP}(B^0\rightarrow K^+\pi^-)$ and 
$A_{CP}(B_s^0\rightarrow \pi^+K^-)$. The strategy used to optimise the
cuts is divided into two steps.
In the first step we define the kinematic cuts against the
combinatorial background, selecting in
an inclusive way the $H_b\rightarrow h^+ h'^-$ candidates, without
using any PID information and assigning by default the pion-mass
hypothesis to all charged tracks. The two
kinematic selections use the same set of cuts, but with different
thresholds. They select pairs of oppositely charged tracks with high 
transverse momentum and large impact parameter with respect to all 
reconstructed primary vertices, fitted in a common vertex displaced
from the related primary vertex.\\
In the second step, exploiting the capabilities of the two RICH
detectors, two sets of PID cuts are defined
(one for each set of optimised kinematic cuts) in order
to separate the data into
%disentangle the various $H_b\rightarrow h^+h'^-$ decay modes into
eight mutually exclusive sub-samples corresponding to distinct final
state hypothesis ($K^+\pi^-$, $K^-\pi^+$, $\pi^+\pi^-$,
$K^+K^-$, $p\pi^-$, $\bar{p}\pi^+$, $p K^-$ and $\bar{p} K^+$). The
guiding principle to identify the PID selection criteria is
to limit the total amount of cross-feed backgrounds under the
$B^0\rightarrow K^+\pi^-$ and $B_s^0\rightarrow\pi^+K^-$ mass peaks to
the same level as the corresponding combinatorial background. Such
cross-feed backgrounds are due to the other $H_b\rightarrow h^+h'^-$
where we mis-identified one or both final state particles. 

\section{Calibration of particle indentification}
The calibration of the PID observables is a crucial aspect of
this analysis, as it is the only variable allowing us to discriminate
between the various decay modes. Hence, in order to determine the
amount of cross-feed backgrounds for a given channel, the
relative efficiencies of the PID selection cuts, employed to identify
the specific final state of interest, play a key role.\\
%The PID variable employed to discriminate between different particle
%hypothesis in LHCb is denoted as $\Delta\log\mathcal{L}_{ij} =
%\Delta\log\mathcal{L}_{i}-\Delta\log\mathcal{L}_{j}$, where
%$\mathcal{L}_i$ and $\mathcal{L}_j$ are the likelihoods for particle
%hypotheses $i$ and $j$ respectively, and $i$ and $j$ stand for 
%charged $\pi$, $K$ or $p$. In order to separate the three particle
%hypothesis we used contemporaneously two $\Delta\log\mathcal{L}_{ij}$
%for each hypothesis: $\Delta\log\mathcal{L}_{K-\pi}$ and
%$\Delta\log\mathcal{L}_{p-\pi}$ for $\pi$, $\Delta\log\mathcal{L}_{K-\pi}$ and
%$\Delta\log\mathcal{L}_{K-p}$ for $K$, $\Delta\log\mathcal{L}_{p-\pi}$ and
%$\Delta\log\mathcal{L}_{K-p}$ for protons.\\
Thanks to the high production rate of $D^*$ mesons at LHC and to the
kinematic characteristics of $D^{*+}\rightarrow D^0(K^-\pi^+)\pi^+$ decay chain (and
its charge conjugate), samples of large statistcs and high purity of $\pi$ and $K$ can be
extracted from these events without any use of PID information. The same
consideration holds for protons obtained from $\Lambda\rightarrow
p\pi^-$ decays.\\
Since production and decay kinematics of $D^0\rightarrow K^-\pi^+$ and
$\Lambda\rightarrow p\pi^-$ channels differ from those of
$H_b\rightarrow h^+h'^-$, the distributions of PID observables 
are reweighted in momentum $p$ and transverse momentum $p_T$, in
order to match the corresponding distributions of particles from
two-body $B$-hadron decays. %As the two PID observables used to
%define each particle hypothesis result to be correlated, their
%simpultaneous calibration is required. 
The efficiencies for each set of PID cuts are evaluated 
from the reweighted distributions.\\

\section{Fits to the $H_b\rightarrow h^+h'^-$ mass spectra}
\begin{center}
\begin{figure}
  
    \rule{0.5cm}{0mm}
    %\hfill\rule{5cm}{0.2mm}
    % \vskip 2.5cm
    %\rule{5cm}{0.2mm}\hfill\rule{5cm}{0.2mm}
    \psfig{figure=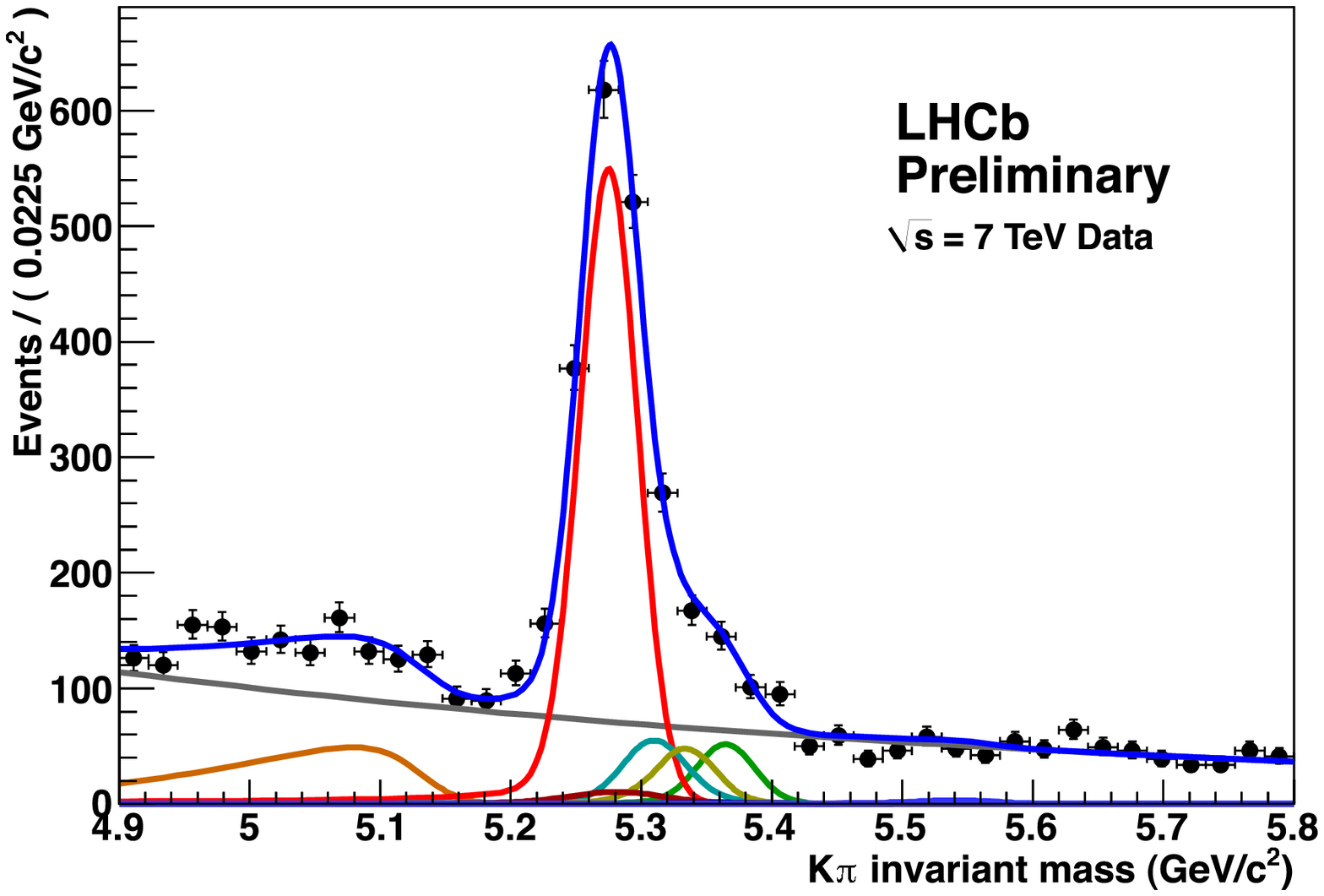,height=2in}
    \psfig{figure=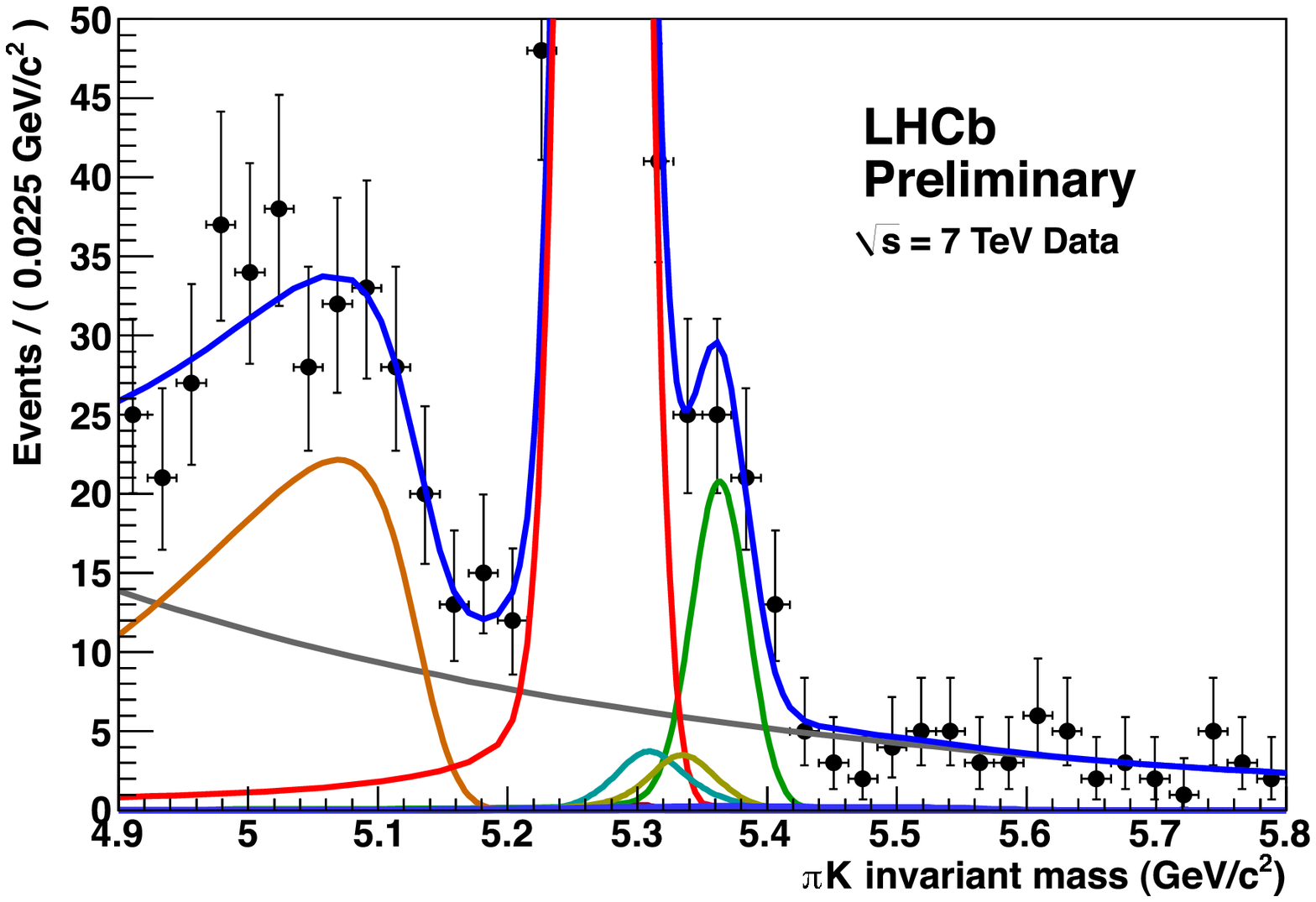,height=2in}
    \caption{$K^+\pi^-$ (plus charge conjugate) invariant mass
      spectrum for events surviving the event selection optimised for
      the best sensitivity on $A_{CP}(B^0\rightarrow K^+\pi^-)$ (left)
      and $A_{CP}(B^0\rightarrow K^+\pi^-)$ (right). The result of the
      unbinned maximum likelihood fit is superimposed. The main
      components contributing to the fit model explained in the text
      are also visible: $B^0\rightarrow K\pi$ (red), wrong sign
      $B^0\rightarrow K \pi$ combination (dark red), $B^0\rightarrow
      \pi^+\pi^-$ (light blue), $B_s^0\rightarrow K^+ K^-$ (dark
      yellow), $B_s^0\rightarrow \pi K$ (green), combinatorial
      background (grey), 3-body partially reconstructed decays (orange).
      \label{fig:fit}}
    
\end{figure}
\end{center}

We perform unbinned maximum likelihood fits to the mass spectra of
events passing the optimised offline selections for
the measurements of $A_{CP}(B^0\rightarrow K^+\pi^-)$ or
$A_{CP}(B_s^0\rightarrow K^-\pi^+)$. The fits are performed
simultaneously on all the eight categories defined by means of the
PID selection criteria.
%The fitting model to the 
%mass spectra takes into account four components. 
The signals, identified as the channels where both tracks are identified with the right mass
hypothesis, are parameterized with a single Gaussian function convolved with a
component accounting for final state QED radiation~\cite{isidori}. The
combinatorial background is modeled with an exponential function. The
invariant mass shapes of cross-feed backgrounds are parameterized by means of
full simulated events, while the normalization of each
mis-identified channel is determined multiplying the yield obtained from the right mass
hypothesis fit by the ratio between PID efficiencies for
the wrong and right final state hypothesis. For the $K^{\pm}\pi^{\mp}$
and $\pi^+\pi^-$ categories it is necessary to model also a
component due to partially reconstructed 3-body $B$-hadron decays, while in
the other final state categories such a contribution is found to be negligible.\\
The results of the fits superimposed to the $K^{\pm}\pi^{\mp}$ mass spectra
(seperately for the samples obtained using the two optimised selections)
are shown in Fig.~\ref{fig:fit}. The asymmetries obtained from the
fits are respectively: $A^{RAW}_{CP}(B^0\rightarrow K^+\pi^-) =
−0.086 \pm 0.033\mathrm{(stat.)}$ and $A^{RAW}_{CP}(B_s^0\rightarrow K^-\pi^+) = 0.15 \pm 0.19\mathrm{(stat.)}$.
The systematic errors due to the fit model and PID calibration are
estimated to be respectively $0.002$ and
$0.004$ for $A_{CP}(B^0\rightarrow K^+\pi^-)$ and $0.021$ and $0.001$ for
$A_{CP}(B_s^0\rightarrow K^+\pi^-)$. 
\section{Correction to the $A_{CP}^{RAW}$}
%The systematic errors on $A_{CP}(B^0\rightarrow K^+\pi^-)$ and
%$A_{CP}(B_s^0\rightarrow K^+\pi^-)$ that we identified can be divided
%into three main category: PID calibration, invariant mass shape model,
%instrumental and production asymmetries.\\
%Systematics related to the PID calibration can affect the $CP$
%asymmentries by an imperfect estimate of the yield of cross-feed
%backgrounds. To estimate the error on the relative PID efficiencies we
%first compared the 
The physical $\mathcal{CP}$ asymmetries we want to measure
are related to the raw asymmetries obtained from
the invariant mass fit by:
\begin{equation}\label{eq:correction}
A_{CP} = A_{CP}^{RAW} - A_{D}(K\pi)-\kappa A_P
\end{equation}
where $A_D(K\pi)$ is the detector induced asymmetry in reconstructing
$K^+\pi^-$ and $K^-\pi^+$ final states, $A_P$ is the production
asymmetry of $B$ mesons and  $\kappa$
is a factor that takes into account the $B-\bar{B}$ oscillation.
The production asymmetry is defined in terms of the $B$ and $\bar{B}$
production rates $A_P= (R_{\bar{B}}-R_B)/(R_{\bar{B}}+R_B)$. The $\kappa$
factor is given by
\begin{equation}
\kappa = \frac{\int (e^{-\Gamma t'}\cos\Delta mt') \varepsilon (t)
  dt}{\int (e^{-\Gamma t'} \cosh \frac{\Delta\Gamma}{2}t') \varepsilon (t) dt},
\end{equation}
where $\varepsilon (t)$ is the acceptance for the decay of interest,
as function of the proper decay time $t$.\\
The detector induced asymmetry $A_D(K\pi)$ is determined using
high statistics samples of tagged $D^{*+}\rightarrow
D^{0} ( K^{-}\pi^{+} ) \pi^{+}$, $D^{*+}\rightarrow D^{0} ( K^{+} K^{-} ) \pi^{+}$ and 
$D^{*+}\rightarrow D^{0} ( \pi^{+} \pi^{-} ) \pi^{+}$, and untagged 
$D^0\rightarrow K^-\pi^+$ decays (plus their charge conjugates). Combining the
integrated raw asymmetries obtained from the invariant mass fit of all
these decay modes and employing the current world average of the
integrated $\mathcal{CP}$ asymmetries for the two modes $D^0\rightarrow K^+K^-$
and $D^0\rightarrow \pi^+\pi^-$~\cite{hfagD}, we determine $A_D(K\pi) = -0.004 \pm
0.004$ (the direct $\mathcal{CP}$ asymmetry of $D^0\rightarrow
K^-\pi^+$ is considered negligible). \\
The production asymmetry $A_P$ is determined by means
of a reconstructed sample of $B^{\pm}\rightarrow J/\psi (\mu^+\mu^-)K^{\pm}$ decays. 
Correcting the raw asymmetry measured from data by the current world average of the direct $\mathcal{CP}$ asymmetry 
$A_{CP}(B^+\rightarrow J/\psi K^+) = 0.009 \pm 0.008$~\cite{pdg}, and
taking into account the reconstruction asymmetry between $K^+$ and
$K^-$ we measure $A_P(B^+) = -0.024 \pm 0.013$. We assume $A_P(B^+)$
equal to $A_P(B^0)$, but introducing a systematic error of $0.01$ to
account for possible differences, obtaining $A_P(B^0) = -0.024 \pm
0.013 \pm 0.010$. Such a systematic uncertainty has been determined by studying the predictions of
different fragmentation models~\cite{LAMBERT}.\\
For the evaluation of the $\kappa$ factors, $\varepsilon (t)$
is determined from full simulated events using the
selections optimised for the respective $A_{CP}(B^0\rightarrow K^+\pi^-)$ and
$A_{CP}(B_s^0\rightarrow K^+\pi^-)$ measurements. The
values of the parameters controlling the time evolution of neutral $B$ mesons, namely $\Gamma_d$, $\Gamma_s$, $\Delta
m_d$, $\Delta m_s$ and $\Delta\Gamma_s$, are taken from the current world
averages~\cite{pdg}, but assuming $\Delta\Gamma_d = 0$. The $\kappa$
factors, computed respectively for the $B^0$ and $B_s^0$, are $\kappa_d
= 0.33$ and $\kappa_s = 0.015$. For the case of the $B_s^0\rightarrow
K^-\pi^+$ decay, even assuming conservatively $A_P(B_s^0) = A_P(B^0)$,
the correction to the $A_{CP}^{RAW}(B_s^0\rightarrow K^-\pi^+)$ results
to be negligible.\\
Using Eq.~\eqref{eq:correction} the central values of the direct
$\mathcal{CP}$ asymmetries are $A_{CP}(B^0\rightarrow K^+\pi^-)=-0.074$ and
$A_{CP}(B_s^0\rightarrow K^+\pi^-)=0.15$.
The statistical errors of $A_D ( K \pi )$ and $\kappa A_P$ are
considered as systematic uncertainties contributing to
$A_{CP}(B^0\rightarrow K^+\pi^-)$
and  $A_{CP}(B_s^0\rightarrow K^+\pi^-)$.

\section{Final result}
Using data collected by the LHCb detector during the 2010 run we
provide preliminary values of the direct CP asymmetries:
\begin{eqnarray}
A_{CP}( B^0\rightarrow K^+\pi^-) = -0.074 \pm 0.033\mathrm{(stat.)} \pm 0.008 \mathrm{(syst.)}\\
A_{CP} (B_s^0 \rightarrow \pi^+K^-) = 0.15 \pm 0.19\mathrm{(stat.)}\pm 0.02\mathrm{(syst.)}.
\end{eqnarray}
The current HFAG average~\cite{hfagB} $A_{CP} (B^0\rightarrow K^+\pi^-) =
-0.098^{+0.012}_{-0.11}$ and the CDF measurement~\cite{cdf} $A_{CP}
(B_s^0\rightarrow K^-\pi^+) = 0.39\pm 0.15\mathrm{(stat.)}\pm
0.08\mathrm{(syst.)}$ are in agreement with our values.

\end{document}